\documentclass[prl,twocolumn,showpacs,preprintnumbers,amsmath,amssymb,superscriptaddress]{revtex4-1}
\usepackage{graphicx}
\usepackage{dcolumn}
\usepackage{bm}
\usepackage{color}
\usepackage{hyperref}
\usepackage{xfrac}
\hypersetup{
 colorlinks,
 linkcolor={red},
 citecolor={blue},
 urlcolor={blue}
}
 \usepackage[ulem=normalem]{changes}

\begin{document}
\preprint{APS/123-QED}

\title{Generic features of the neutron-proton interaction}


\author{Y.~H.~Kim}
\affiliation{GANIL, CEA/DRF-CNRS/IN2P3, Bd Henri Becquerel, BP 55027, F-14076 Caen Cedex 5, France}
\email{yunghee.kim@ganil.fr}
\author{M.~Rejmund}
\affiliation{GANIL, CEA/DRF-CNRS/IN2P3, Bd Henri Becquerel, BP 55027, F-14076 Caen Cedex 5, France}
\author{P.~Van~Isacker}
\affiliation{GANIL, CEA/DRF-CNRS/IN2P3, Bd Henri Becquerel, BP 55027, F-14076 Caen Cedex 5, France}
\author{A.~Lemasson}
\affiliation{GANIL, CEA/DRF-CNRS/IN2P3, Bd Henri Becquerel, BP 55027, F-14076 Caen Cedex 5, France}

\date{\today} 

\begin{abstract}
We show that fully aligned neutron-proton pairs
play a crucial role in the low-energy spectroscopy of nuclei.
 with valence nucleons in a high-$j$ orbital.
Their dominance is valid in nuclei with valence neutrons and protons
in different high-$j$ orbitals as well as in $N=Z$ nuclei,
where all nucleons occupy the same orbital.
We demonstrate analytically this generic feature of the neutron-proton interaction
for a variety of systems with four valence nucleons
interacting through realistic, effective forces.
The dominance of fully aligned neutron-proton pairs results from the combined effect of
(i) angular momentum coupling and
(ii) basic properties of the neutron-proton interaction.
\end{abstract}
\pacs{} 

\maketitle 

Insight in the properties of correlated quantum many-body systems
often can be obtained by means of solvable models.
In condensed-matter physics arguably the most important example
is the Hubbard model, for which an exact solution is known in one dimension~\cite{Lieb68};
many other examples are reviewed in Ref.~\cite{Dukelsky04}. 
In nuclear physics, solvable models fall into two broad classes,
depending on whether the nucleons interact through a pairing or a quadrupole force.
The former case, first considered by Racah in the context of atomic physics~\cite{Racah43},
applies to `semi-magic' nuclei with either neutrons or protons in the valence shell,
and leads to a classification of states in terms of seniority
or, equivalently, in terms of the number of unpaired nucleons.
The quadrupole force, on the other hand,
is appropriate for nuclei with several neutrons {\em and} protons in the valence shell,
and gives rise to a rotational classification of states based on SU(3) symmetry~\cite{Elliott58}.
While their solution may be exact,
such models provide at best approximations to observed quantum many-body systems.

Some time ago Cederwall {\it et al.}~\cite{Cederwall10}
suggested that in self-conjugate nuclei
(i.e., $N=Z$ nuclei with equal numbers of neutrons and protons)
yet a different classification scheme may exist.
Although not based on the property of solvability,
their analysis claimed an analogy with pairing:
While in traditional pairing models states are analyzed in terms of $J=0$ pairs
composed of either two neutrons ($\nu\nu$) or two protons ($\pi\pi$),
the approach of Cederwall {\it et al.} centered on neutron-proton ($\nu\pi$) pairs
that are fully aligned in angular momentum,
which we refer to as fully-aligned $\nu\pi$ pairs (FAPs). 
The claim for the existence of this novel coupling scheme
was based on the measured spectrum of $^{92}_{46}$Pd$^{\phantom{00}}_{46}$
and backed by a theoretical analysis in the framework of the nuclear shell model~\cite{Qi11,Xu12}. 
The idea of FAPs
(which also relates to the stretch scheme proposed by Danos and Gillet~\cite{Danos67})
provoked a flurry of studies based on boson mappings~\cite{Zerguine11,Isacker13}
or symmetry techniques~\cite{Neergard13,Neergard14},
or with use of schematic interactions~\cite{Zamick13,Fu13c,Fu14} in a single-$j$ approach.
Simultaneously, also multi-$j$ calculations with realistic interactions
were carried out~\cite{Coraggio12,Fu13b,Fu15,Sambataro15,Robinson16un}
to test the prevalence of FAPs in $N=Z$ nuclei. 
A review can be found in Ref.~\cite{Frauendorf14}. 
Despite occasional contradictions,
a general consensus seems to emerge that FAPs
are dominant (with some caveats---see below)
if neutrons and protons occupy a single-$j$ orbital
but that their dominance fades away quickly in a multi-$j$ scenario.
In addition, it has also become clear
that the different proposed schemes are not mutually exclusive
because of their non-orthogonality,
and that, for example, nuclear ground states
can be equally well described
in terms of $J=0$ and $J=1$ pairs~\cite{Sambataro15}.

It should nevertheless be stressed
that studies of the FAP conjecture so far have been limited to $N=Z$ nuclei.
In this Rapid Communication we point out that the FAP dominance
is a widely occurring phenomenon,
not confined to the restricted class of $N=Z$ nuclei,
but present throughout the nuclear chart
whenever valence neutrons and protons mainly occupy high-$j$ orbitals.
We show that the FAP dominance results from the combined effect of
(i) angular momentum coupling and
(ii) some basic properties of the $\nu\pi$ interaction.

We consider nuclei with neutrons and protons in valence orbitals $j_\nu$ and $j_\pi$, respectively,
and concentrate on nuclei with two neutrons and two protons ($2+2$),
briefly exploring at the end the case ($4+4$).
The relevant $\nu\nu$, $\pi\pi$, and $\nu\pi$
two-body matrix-elements (TBMEs) of the residual interaction
are denoted by $V_{\nu\nu}^{J_\nu}$, $V_{\pi\pi}^{J_\pi}$, and $V_{\nu\pi}^{J_{\nu\pi}}$, respectively,
where $J_\nu$, $J_\pi$, and $J_{\nu\pi}$ are the coupled angular momenta of two nucleons.
For a nucleus with two neutrons and two protons in the valence orbitals,
the matrix element of the Hamiltonian
in a $\nu\nu$-$\pi\pi$ basis with vectors $|J_\nu J_\pi;J\rangle$
can be expressed in terms of the TBMEs as follows:
\begin{align}
&\langle J_\nu J_\pi;J|\hat H|J'_\nu J'_\pi;J\rangle
\nonumber\\&\quad=
(V_{\nu\nu}^{J_\nu}+V_{\pi\pi}^{J_\pi})\delta_{J_\nu J'_\nu}\delta_{J_\pi J'_\pi} 
+\sum_{J_{\nu\pi}}C_{J_{\nu\pi}}V_{\nu\pi}^{J_{\nu\pi}},
\label{eq:me} 
\end{align}
where $C_{J_{\nu\pi}}\equiv C_{J_{\nu\pi}}(j_\nu,j_\pi,J_\nu,J_\pi,J'_\nu,J'_\pi,J)$
is an angular-momentum coupling coefficient,
whose explicit expression is given in the Supplemental Material.
Note that the $\nu\nu$ and $\pi\pi$ interactions
contribute only to Hamiltonian matrix elements that are diagonal,
and that solely the $\nu\pi$ interaction
is responsible for the mixing in the $\nu\nu$-$\pi\pi$ basis.

\begin{figure}
\includegraphics[width=0.85\columnwidth]{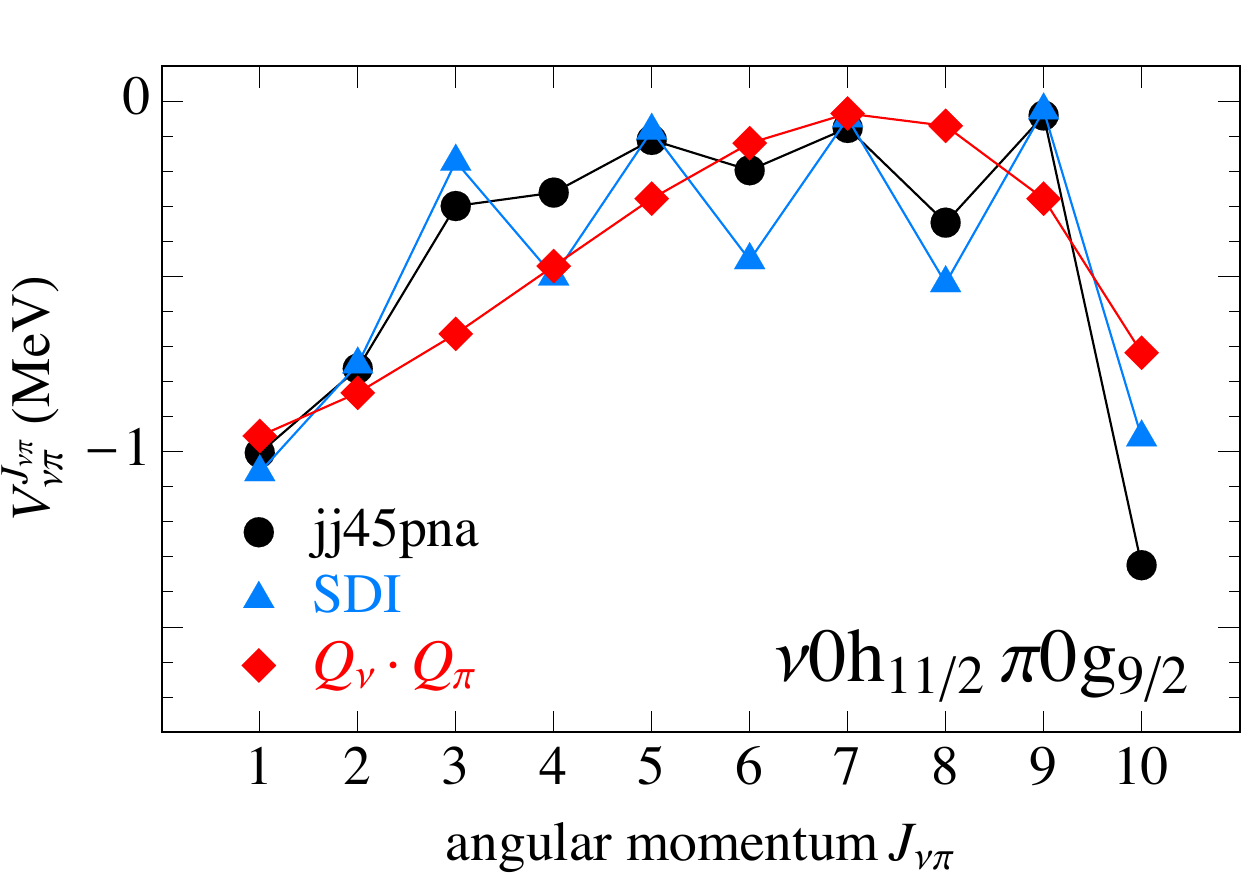}
\caption{(color online) 
The TBMEs for three different $\nu0h_{11/2}\pi0g_{9/2}$ interactions:
(i) realistic (jj45pna, black circles),
(ii) surface delta (SDI, blue triangles),
and (iii) quadrupole ($Q_\nu\cdot Q_\pi$, red squares).}
\label{fig:vnp3} 
\end{figure}
In the following we present a detailed discussion of $^{128}$Cd,
which has two proton and two neutron holes with respect to the closed-shell nucleus $^{132}$Sn
and we assume that the low-energy states
are dominated by the high-$j$ intruder orbitals $\nu0h_{11/2}$ and $\pi0g_{9/2}$.
This assumption can be verified
in a large-scale shell-model calculation with a realistic residual interaction
and a valence space consisting of $\nu 1d_{3/2}, 2s_{1/2}, 0h_{11/2}$
and $\pi 1p_{1/2}, 0g_{9/2}$~\cite{Rejmund16}.
For the positive-parity yrast states with $J^\pi$ from $0^+$ to $18^+$,
the calculated occupation probabilities of $\nu 0h_{11/2}$ and $\pi 0g_{9/2}$
are $\gtrsim$10 and $\gtrsim$8, respectively,
which strongly supports our assumption.
The $\nu\nu$ and $\pi\pi$ TBMEs are set so as to reproduce
the experimental spectra of $^{130}$Sn and $^{130}$Cd, respectively.
To test the robustness of our hypothesis concerning the FAP dominance,
we use three different $\nu\pi$ TBMEs shown in Fig.~\ref{fig:vnp3}:
(i) the realistic interaction jj45pna~\cite{Hjorth-Jensen95},
(ii) a surface delta interaction (SDI) with isoscalar and isovector strengths adjusted to jj45pna,
and (iii) a quadrupole interaction $Q_\nu\cdot Q_\pi$ (plus a constant).
The latter schematic interaction is included here
to study its relation to the FAP scheme~\cite{Zuker15}.

\begin{figure}
\includegraphics[width=\columnwidth]{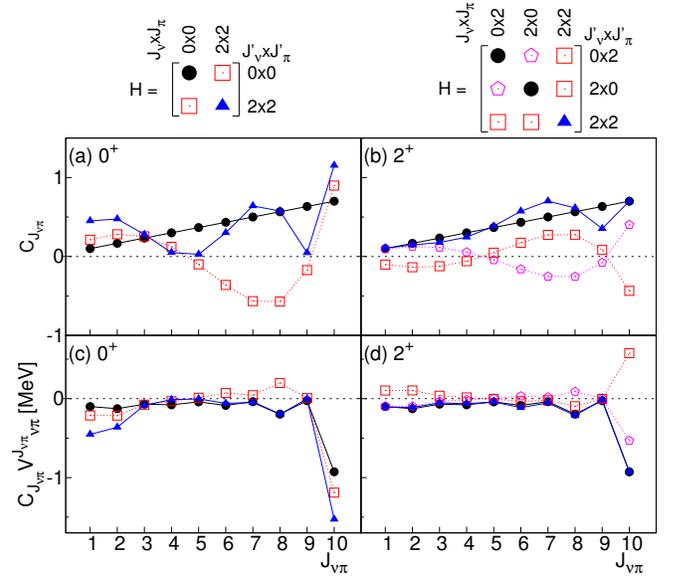}
\caption{(color online) 
Panels (a) and (b):
The coefficient $C_{J_{\nu\pi}}$ in Eq.~(\ref{eq:me}) for $j_\nu=11/2$ and $j_\pi=9/2$
as a function of $J_{\nu\pi}$ for $J=0$ and $J=2$.
Panels (c) and (d):
The energy contribution in the $\nu\nu$-$\pi\pi$ basis
due to the individual TBMEs of the jj45pna interaction
as a function of $J_{\nu\pi}$ for $J=0$ and $J=2$.
Different symbols correspond to various couplings $(J_\nu,J_\pi,J'_\nu,J'_\pi)$ as indicated. 
Full (open) symbols indicate diagonal (off-diagonal) matrix elements in the $\nu\nu$-$\pi\pi$ basis.}
\label{fig:ccoef} 
\end{figure}
In Fig.~\ref{fig:ccoef} the coefficient $C_{J_{\nu\pi}}$ in Eq.~(\ref{eq:me})
is shown as a function of $J_{\nu\pi}$ for $J=0$ and $J=2$. 
This coefficient strongly suppresses the contribution of low-$J_{\nu\pi}$ interactions
and favors the interaction in the state with angular momentum $J_{\nu\pi}^{\rm FAP}\equiv j_\nu+j_\pi$.
This is so in particular for $J=J_\nu=J_\pi=J'_\nu=J'_\pi=0$,
in which case $C_{J_{\nu\pi}}$ is proportional to $2J_{\nu\pi}+1$.
The contribution of the FAP is strongest also for the off-diagonal matrix elements
and therefore dominates the mixing in the $\nu\nu$-$\pi\pi$ basis.
This is further illustrated in Fig.~\ref{fig:ccoef}
by showing the energy contribution in the $\nu\nu$-$\pi\pi$ basis
due to the jj45pna TBMEs as a function of $J_{\nu\pi}$ for $J=0$ and $J=2$. 
The only significant contribution is observed for the FAP
while the low-$J_{\nu\pi}$ interactions represent a nearly-constant, small contribution.

\begin{figure}
\includegraphics[width=\columnwidth]{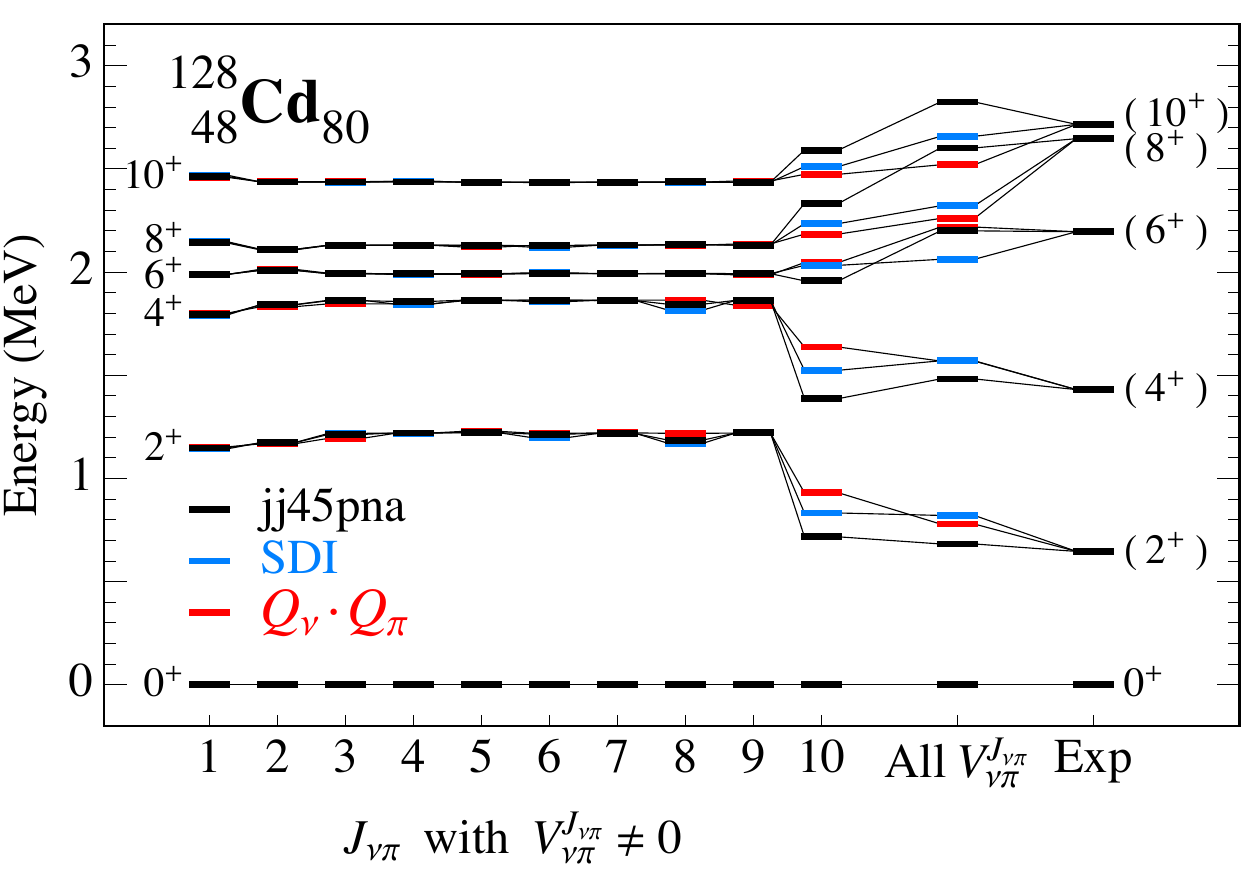}
\caption{(color online) 
Energies of the yrast eigenstates of the Hamiltonian~(\ref{eq:me})
for two neutrons and two protons ($^{128}$Cd)
as a function of $J_{\nu\pi}$,
the angular momentum of the only non-zero component
in the $\nu\pi$ interaction ($V_{\nu\pi}^{J_{\nu\pi}} \neq 0$).
The results are for three different $\nu\pi$ interactions:
(i) realistic (jj45pna, black),
(ii) surface delta (SDI, blue),
and (iii) quadrupole ($Q_\nu\cdot Q_\pi$, red).
Also the eigenenergies of the full Hamiltonian
and the experimental spectrum of $^{128}$Cd are shown.}
\label{fig:cd128e3} 
\end{figure}
In Fig.~\ref{fig:cd128e3} are shown the yrast levels resulting
from the diagonalization of the Hamiltonian matrix~(\ref{eq:me})
with the full $\nu\nu$ and $\pi\pi$ interactions
but with only a single non-zero component $J_{\nu\pi}$
($V_{\nu\pi}^{J\neq J_{\nu\pi}}=0$) of the above-mentioned $\nu\pi$ interactions.
The levels are shown as a function of $J_{\nu\pi}$
and compared to those obtained with the full Hamiltonian
as well as to the experimental results. 
Clearly, the interaction in the FAP state
has the crucial impact on the level energies,
and this is so for the three interactions.
The close agreement between observed level energies
and those calculated with the jj45pna interaction
further justifies the assumption that the structure of the low-energy states
is dominated by the high-$j$ intruder orbitals $\nu0h_{11/2}$ and $\pi0g_{9/2}$.
The correspondence between the spectra
calculated with a single non-zero component $J_{\nu\pi}=10$
and with the full Hamiltonian is remarkable for jj45pna and SDI.
The FAP dominance is still observed for the quadrupole interaction
albeit to a lesser extent.
With reference to Fig.~\ref{fig:vnp3} it is clear that the driving force
behind the mechanism of the FAP dominance
is the TBME for $J_{\nu\pi}=j_\nu+j_\pi$
since for the schematic quadrupole interaction 
this matrix element is less attractive than it is for jj45pna or SDI.

The dominant components of the $0^+_1$ and $2^+_1$ states
resulting from the complete jj45pna calculation are
\begin{align}
|0^+_1\rangle&\approx0.82|0_\nu0_\pi;0\rangle+0.52|2_\nu2_\pi;0\rangle+\cdots,
\nonumber\\
|2^+_1\rangle&\approx0.62|2_\nu0_\pi;2\rangle+0.56|0_\nu2_\pi;2\rangle-0.36|2_\nu2_\pi;2\rangle+\cdots,
\nonumber
\end{align}
while those obtained with $V_{\nu\pi}^{10^-}$ only read
\begin{align}
|0^+_1\rangle&\approx0.89|0_\nu0_\pi;0\rangle+0.43|2_\nu2_\pi;0\rangle+\cdots,
\nonumber\\
|2^+_1\rangle&\approx0.66|2_\nu0_\pi;2\rangle+0.60|0_\nu2_\pi;2\rangle-0.34|2_\nu2_\pi;2\rangle+\cdots.
\nonumber
\end{align}
Note the important $|2_\nu2_\pi;0\rangle$ component in the $0^+$ ground state
as well as the qualitative agreement of the wave functions in both calculations.

\begin{figure}
\includegraphics[width=0.9\columnwidth]{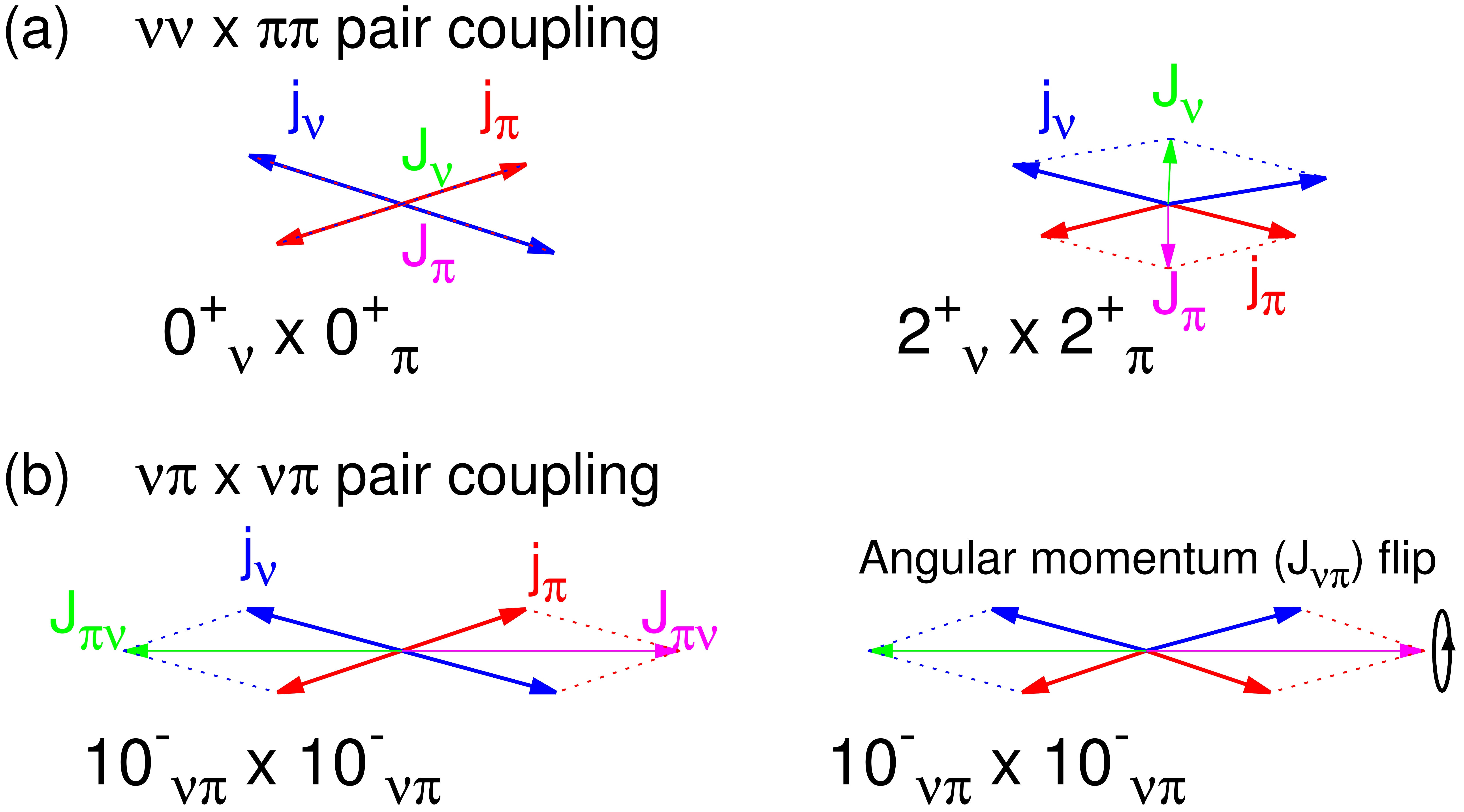}
\caption{(color online)
Graphical illustration of the (a) $\nu\nu$-$\pi\pi$
and (b) $\nu\pi$-$\nu\pi$ basis for a $0^+$ state.
The dashed lines indicate the angular momentum vector adding scheme for different pairs.
In (b) both couplings are identical
but one pair is spin-flipped around its total angular momentum axis.}
\label{fig:npcoupling} 
\end{figure}
So far we considered a pair of neutrons coupled to a pair of protons
and found considerable mixing in the $\nu\nu$-$\pi\pi$ basis
due to the $\nu\pi$ interaction.
Alternatively, $\nu\pi$ pairs can be considered
for the construction of a $\nu\pi$-$\nu\pi$ basis with vectors $|J^1_{\nu\pi}J^2_{\nu\pi};J\rangle$
(see Supplemental Material).
It is important to realize that the $\nu\nu$-$\pi\pi$ basis is orthogonal
whereas the $\nu\pi$-$\nu\pi$ basis is not.
Also, the states in both bases may have large overlaps,
depending on the coupling of the angular momenta.
This is illustrated for the $0^+$ state in Fig.~\ref{fig:npcoupling},
where states in both bases are represented graphically,
leading to similar or dissimilar configurations, as the case may be.

\begin{figure}
\includegraphics[width=0.8\columnwidth]{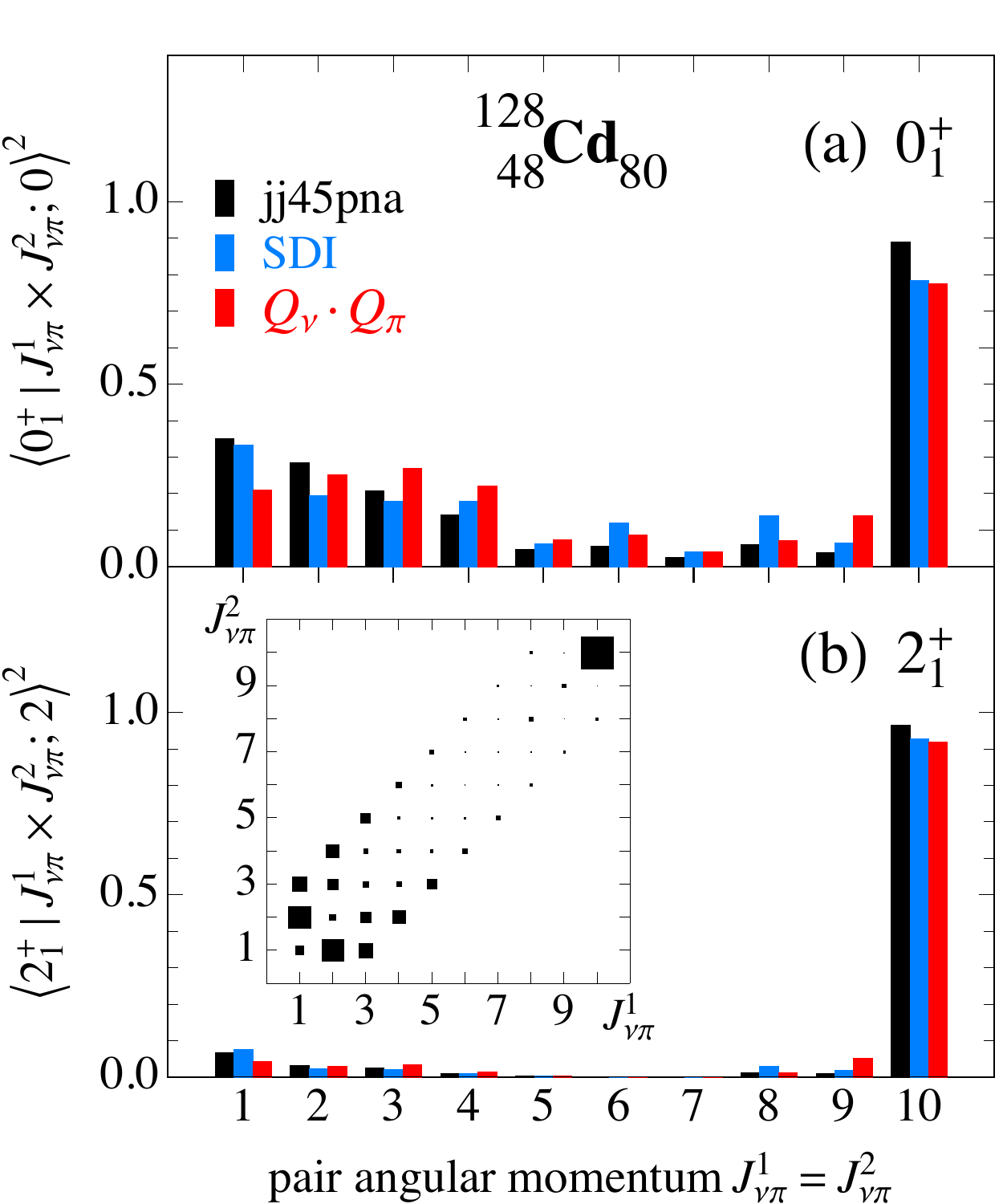}
\caption{(color online) 
Squares of the components of the yrast eigenstates of $^{128}$Cd
expressed in the $\nu\pi$-$\nu\pi$ basis $|J^1_{\nu\pi}J^2_{\nu\pi};J\rangle$
for (a) $J=0$ and (b) $J=2$.
The main figure shows $\langle J_1|J^1_{\nu\pi}J^2_{\nu\pi};J\rangle^2$
as a function of $J^1_{\nu\pi}=J^2_{\nu\pi}$
for three different $\nu\pi$ interactions:
(i) realistic (jj45pna, black),
(ii) surface delta (SDI, blue),
and (iii) quadrupole ($Q_\nu\cdot Q_\pi$, red).
In the inset of (b) this quantity is shown (for jj45pna)
as proportional to the size of the squares
as a function of $J^1_{\nu\pi}$ and $J^2_{\nu\pi}$.}
\label{fig:cd128wf} 
\end{figure}
Figure~\ref{fig:cd128wf} displays, for the three different $\nu\pi$ interactions,
the squares of the components of the yrast eigenstates of the full Hamiltonian for $^{128}$Cd,
expressed in the $\nu\pi$-$\nu\pi$ basis $|J^1_{\nu\pi}J^2_{\nu\pi};J\rangle$ for $J=0$ and $J=2$.
We emphasize that the $\nu\pi$-$\nu\pi$ basis is non-orthogonal
and therefore the sum of the squares may exceed 1.
Nevertheless, from Fig.~\ref{fig:cd128wf}
it follows that, independent of the $\nu\pi$ interaction,
the biggest and dominant contribution
to the $0_1^+$ and $2_1^+$ states comes from the FAP
with $J^1_{\nu\pi}=J^2_{\nu\pi}=J^{\rm FAP}_{\nu\pi}$.
The low-$J_{\nu\pi}$ pairs have a non-negligible contribution to $0_1^+$
but contribute little to $2_1^+$.
Further, the inset of Fig.~\ref{fig:cd128wf}(b)
shows that the second strongest contribution to $2_1^+$
results from $(J^1_{\nu\pi},J^2_{\nu\pi})=(1,2)$ [or $(2,1)$].

\begin{figure}
\includegraphics[width=0.8\columnwidth]{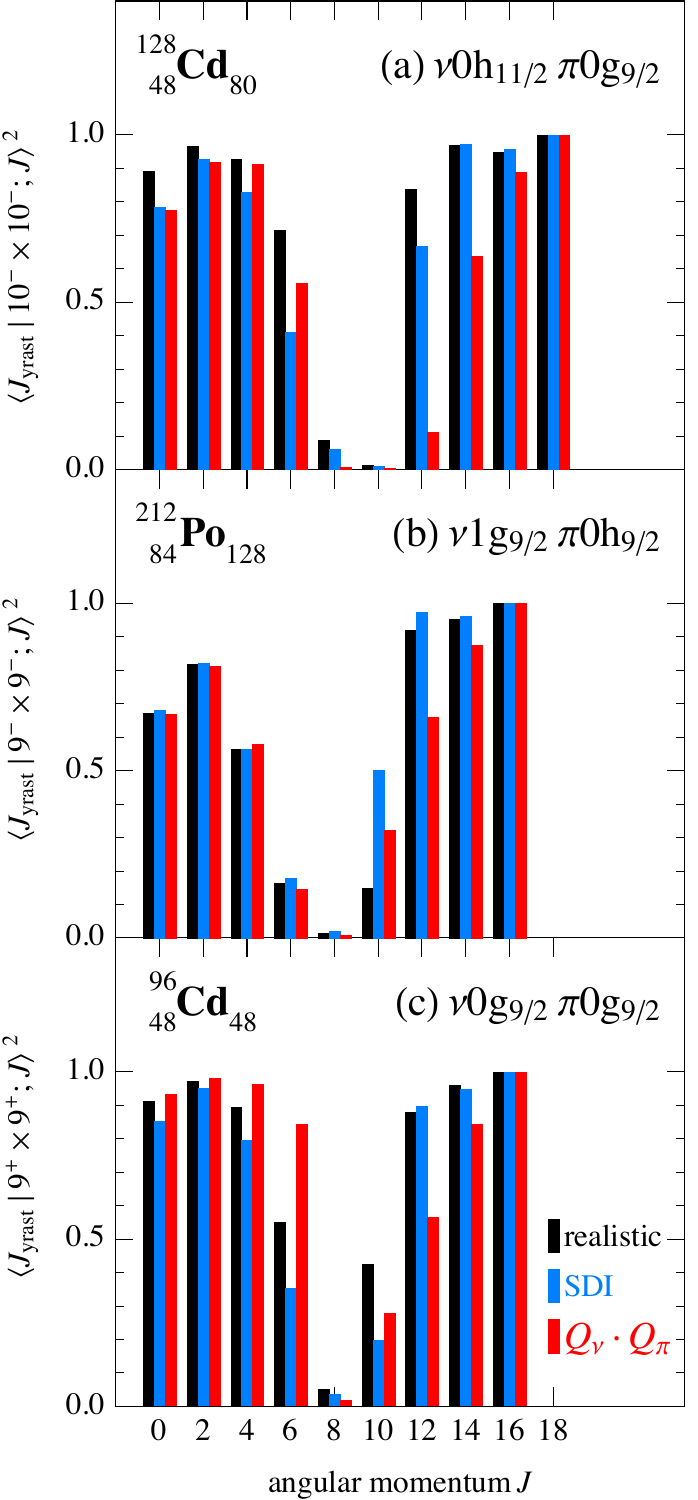}
\caption{(color online)
Overlaps $\langle J_{\rm yrast}|(J^{\rm FAP}_{\nu\pi})^2;J\rangle^2$
as a function of the angular momentum $J$ for
(a) $^{128}$Cd,
(b) $^{212}$Po,
and (c) $^{96}$Cd
calculated with a realistic interaction (black),
and compared to the surface delta (SDI, blue)
and quadrupole ($Q_\nu\cdot Q_\pi$, red) interactions.}
\label{fig:cdpowf} 
\end{figure}
This analysis can be repeated for other yrast eigenstates of $^{128}$Cd,
as is done in Fig.~\ref{fig:cdpowf}(a),
which shows, as a function of the angular momentum $J$ of the state,
the overlaps $|\langle J_{\rm yrast}|(J^{\rm FAP}_{\nu\pi})^2;J\rangle|^2$,
where $|J_{\rm yrast}\rangle$ is an eigenstate of the full Hamiltonian.
Again independent of the $\nu\pi$ interaction,
the FAP is seen to give rise to the major component
of the low- and high-$J$ yrast eigenstates
while its contribution is small at mid $J$.
The $J=8$ and $J=10$ yrast eigenstates
are fragmented in the $\nu\pi$-$\nu\pi$ basis
and their structure is more transparent in the $\nu\nu$-$\pi\pi$ basis,
where they correspond dominantly to a seniority $\upsilon=2$ state,
{\it i.e.} to $|0^+_\nu 8^+_\pi;8^+\rangle$ and $|10^+_\nu 0^+_\pi;10^+\rangle$, respectively.

To demonstrate the generic character of the results for $^{128}$Cd,
a similar analysis can performed for different regions of nuclear chart.
In each case we consider a `realistic' interaction, taken from or fitted to data,
and compare it with the surface delta and quadrupole interactions.
The first example concerns $^{212}$Po,
which has two neutrons and two protons outside the closed-shell nucleus $^{208}$Pb.
Assuming that the neutrons are in $\nu1g_{9/2}$ and the protons in $\pi0h_{9/2}$,
an effective interaction for this model space
can be extracted from the observed spectra of $A=210$ nuclei~\cite{ShamsuzzohaBasunia2014}.
Figure~\ref{fig:cdpowf}(b) shows the overlaps
$\langle J_{\rm yrast}|(J^{\rm FAP}_{\nu\pi})^2;J\rangle^2$
involving the FAP state with $J^{\rm FAP}_{\nu\pi}=9$.
The behavior is similar to that found in $^{128}$Cd
but at low $J$ the FAP dominance is less pronounced.
This illustrates the crucial role of the TBME in the FAP state,
which is less attractive in $^{212}$Po,
owing to the combination of a neutron orbital with $j_\nu=\ell_\nu+{\sfrac12}$
and a proton orbital with $j_\pi=\ell_\pi-{\sfrac12}$  as compared to  $^{128}$Cd
with  $j_\nu=\ell_\nu+{\sfrac12}$ and $j_\pi=\ell_\pi+{\sfrac12}$.
In addition, we present in Fig.~\ref{fig:cdpowf}(c)
an application to an $N=Z$ nucleus, $^{96}$Cd,
where neutrons and protons occupy the same orbital $0g_{9/2}$,
using the interaction from Ref.~\cite{Serduke1976}.
This case again is markedly similar to what is found in $^{128}$Cd.

\begin{figure}
\includegraphics[width=\columnwidth]{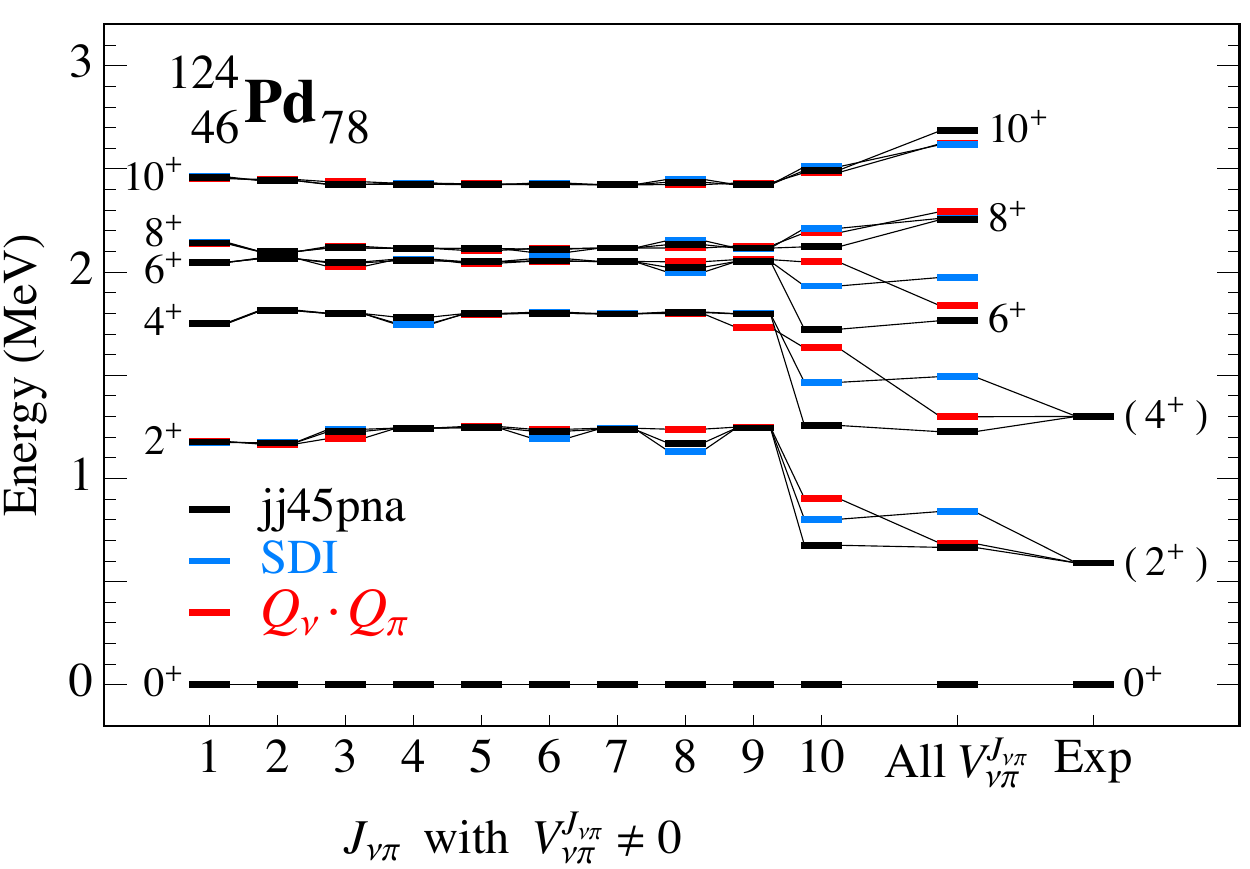}
\caption{(color online) 
Same as Fig.~\ref{fig:cd128e3} but for four neutrons and four protons ($^{124}$Pd).}
\label{fig:pd124e3} 
\end{figure}
Finally, the above analysis can be extended to four neutrons and four protons.
In Fig.~\ref{fig:pd124e3} are shown the results for $^{124}$Pd, 
obtained in the same way as those for $^{128}$Cd in Fig.~\ref{fig:cd128e3}.
Again we find that the results of the calculation with the FAP interaction
are close to those obtained with the complete $\nu\pi$ interaction,
which in turn are in close agreement with experiment~\cite{Wang13}.
This result for $^{124}$Pd can be seen
as the analogue to that for $^{92}$Pd~\cite{Cederwall10}.
The regularly spaced level sequence,
claimed to be the experimental signature
of the dominance of the FAP in $^{92}$Pd~\cite{Cederwall10},
is also observed for $^{124}$Pd.

We conclude therefore that,
if valence nucleons dominantly occupy high-$j$ orbitals,
the yrast spectroscopy of the nucleus
is essentially determined by a single matrix element,
namely the one where neutron and proton are fully aligned in angular momentum.
This property can be considered as a generic feature of the neutron-proton interaction,
akin to pairing between identical nucleons.
Since the seminal papers of Racah and Elliott,
one of the friendly companions of nuclear spectroscopists
has been the pairing-plus-quadrupole model.
With this study we suggest that the mechanism
at the basis of the success of this schematic Hamiltonian
is the interaction in the configuration
with fully aligned neutron and proton angular momenta
and that therefore a study of 
possible symmetries of the pairing-plus-FAP model is called for.


\bibliography{genericnp}

\bibliographystyle{myapsrev4-1}

\end{document}